\documentclass[twocolumn,aps,superscriptaddress,showpacs]{revtex4}
\usepackage{amssymb}
\usepackage{amsmath}
\usepackage{graphicx}
\usepackage[normalem]{ulem}
\usepackage[dvips]{color}

\def\esym{$E_{sym}(\rho)$~}
\begin{document}

\title{Super-soft symmetry energy encountering non-Newtonian gravity in neutron stars}

\author{De-Hua Wen}
\affiliation{Department of Physics, South China University of
Technology,Guangzhou 510641, P.R. China} \affiliation{Department
of Physics and Astronomy, Texas A\&M University-Commerce,
Commerce, Texas 75429-3011, USA}
\author{Bao-An Li\footnote{Corresponding author, Bao-An\_Li$@$Tamu-Commerce.edu}}
\affiliation{Department of Physics and Astronomy, Texas A\&M
University-Commerce, Commerce, Texas 75429-3011, USA}
\author{Lie-Wen Chen}
\affiliation{Department of Physics, Shanghai Jiao Tong University,
Shanghai 200240, P.R.China}
\date{\today}

\begin{abstract}
Considering the non-Newtonian gravity proposed in the grand
unification theories, we show that the stability and observed global
properties of neutron stars can not rule out the super-soft nuclear
symmetry energies at supra-saturation densities. The degree of
possible violation of the Inverse-Square-Law of gravity in neutron
stars is estimated using an Equation of State (EOS) of neutron-rich
nuclear matter consistent with the available terrestrial laboratory
data.
\end{abstract}


 \maketitle

The density dependence of nuclear symmetry energy \esym is an
important ingredient for understanding many interesting phenomena in
astrophysics, cosmology \cite{Sum94,Bom01,Lat04,Ste05a} and nuclear
physics \cite{LiBA98,Dan02a,Bar05,LCK08,Bro00}. However, theoretical
predictions on the \esym especially at supra-saturation densities
are currently rather diverse
\cite{LCK08,Bro00,Kut94,Kub99,Szm06,Dip03}. Unfortunately, there is
no known first-principle guiding the high-density behavior of the
\esym. Presently, while many theories, see, e.g., refs.
\cite{Ste05a,Lee98,Hor01a,Dip03,Che07,LiZH06}, predict that the
\esym increases continuously at all densities, many other models,
see, e.g., refs.
\cite{Pan72,Fri81,Wir88a,eft,Kra06,Bro00,Cha97,Sto03,Che05b,Dec80,Das03,Kho96,Bas07,MS,Ban00,Ch09},
predict that the \esym first increases and then decreases above
certain supra-saturation densities. The \esym may even become
negative at high
densities~\cite{Bom01,Wir88a,LiBA98,LCK08,Bro00,Sto03,Kut94,Kub99,Szm06}.
This latter kind of symmetry energy functions are generally regarded
as being soft. Some (e.g., the UV14+TNI in \cite{Wir88a} and group
II in \cite{Sto03}) of them can describe very well all observed
properties of neutron stars (NSs). However, the super-soft ones
(e.g., the original Gogny-Hartree-Fock (GHF) prediction \cite{Das03}
and group III in \cite{Sto03}) that quickly drops to zero around
three times the saturation density $\rho_0$ either can not keep the
NSs stable or predict maximum NS masses significantly below
$1.4~\textrm{M}_{\odot}$ depending on the EOS used for symmetric
nuclear matter. Given the above theoretical situation, experimental
indications on the high density \esym are thus utmost important.
Very interestingly, circumstantial evidence for a super-soft \esym
\cite{Xiao09} was found very recently from analyzing the FOPI/GSI
experimental data on the $\pi^-/\pi^+$ ratio in relativistic
heavy-ion collisions \cite{Rei07} within a transport model
\cite{IBUU04} using the MDI (Momentum-Dependent-Interaction) EOS
\cite{Das03}. While the symmetric part of the MDI EOS is consistent
with the existing terrestrial nuclear laboratory
data\cite{Dan02a,LCK08}, the total pressure of NS matter obtained
using the super-soft \esym (which is actually the original GHF
prediction) preferred by the FOPI/GSI data can not keep neutron
stars stable. Among possibly many important ramifications in
astrophysics and cosmology, this finding posts immediately a serious
scientific challenge: how can the NSs be stable with such kind of
super-soft symmetry energies? In fact, this question has been raised
repeatedly and the answer has been negative long before any
experimental indication was available. In the literature, the
super-soft symmetry energies were often regarded by some people as
either ``unpleasant", see, e.g., \cite{Cha97}, or ``unphysical",
see, e.g., \cite{Glen,Sto03,Stone05}. These assertions, of course,
are all based on the assumption that gravity is well understood.
However, it is really remarkable that gravity, despite being the
first to be discovered, is actually still considered by far the most
poorly understood force \cite{Pea01,Hoy03,Ark98}. In fact, in
pursuit of unifying gravity with the three other fundamental forces,
conventional understanding about gravity has to be modified due to
either the geometrical effect of the extra space-time dimensions
predicted by string theories and/or the exchange of the weakly
interacting bosons newly proposed in the super-symmetric extension
of the Standard Model, see, e.g., refs. \cite{Fis99,Adel03} for
reviews. Consequently, the Inverse-Square-Law (ISL) of gravity is
expected to be violated. In stable neutron stars at $\beta$
equilibrium which is determined by the weak and electromagnetic
interactions, the gravity has to be balanced by the strong
interaction. Neutron stars are thus a natural testing ground of
grand unification theories. In this Letter, we show that the
super-soft \esym preferred by the FOPI/GSI data can readily keep
neutron stars stable if the non-Newtonian gravity is considered.

The deviation from the ISL of gravity can be characterized
effectively by adding a Yukawa term to the normal gravitational
potential \cite{Fujii71,Long03}, i.e.,
\begin{equation}\label{Vr}
V(r)=-\frac{G_{\infty}m_{1}m_{2}}{r}(1+\alpha e^{-r/\lambda}),
\end{equation}
where $\alpha$ is a dimensionless strength parameter, $\lambda$ is
the length scale and $G_{\infty}$ is the universal gravitational
constant. Alternatively, the Yukawa term can also be considered as
due to the putative ``fifth force" \cite{Fis99,Adel03,Fujii71}
coexisting with gravity or a non-universal gravitational
``constant" \cite{Fis99,Uzan03} of $G(r)=G_{\infty}[1+\alpha
e^{-r/\lambda}(1+r/\lambda)]$. In the scalar/vector boson exchange
picture, $\alpha=\pm g^2/(4\pi G_{\infty}m_b^{2})$ and
$\lambda=1/\mu$ (in natural units). The $g^2$, $\mu$ and $m_b$ are
the boson-baryon coupling constant, the boson and baryon mass,
respectively. To reduce gravity from the ISL, the exchange of a
vector boson is necessary. It is worth noting that a neutral
spin-1 vector $U$-boson has been a favorite candidate. It is very
weakly coupled to baryons \cite{Kri09}, can mediate the
interactions among Dark Matter (DM) candidates \cite{Fayet,Boe04}
and has been used to explain the 511 keV $\gamma$-ray observation
from the galactic bulge \cite{Jean03,Boehm04a,Zhu07}.

According to Fujii \cite{Fuj2}, the Yukawa term is simply part of
the matter system in general relativity. Consequently, the
Einstein equation remains the same and only the EOS is modified.
Within the mean-field approximation, the extra energy density due
to the Yukawa term is \cite{Long03,Kri09}
\begin{equation}\label{EDUB}
\varepsilon_ {_{\textrm{UB}}}= \frac{1}{2V}\int
\rho(\vec{x}_{1})\frac{g^{2}}{4\pi}\frac{e^{-\mu
r}}{r}\rho(\vec{x}_{2})d\vec{x}_{1}d\vec{x}_{2}=\frac{1}{2}\frac{g^{2}}{\mu^{2}}\rho^{2},
\end{equation}
where $V$ is the normalization volume, $\rho$ is the baryon number
density and $r=|\vec{x}_{1}-\vec{x}_{2}|$. The corresponding
addition to the pressure is then $
P_{UB}=\frac{1}{2}\frac{g^{2}\rho^2}{\mu^{2}}\left(1-\frac{2\rho}{\mu}\frac{\partial
\mu}{\partial \rho}\right). $ Assuming a constant boson mass
independent of the density, one obtains
$P_{UB}=\varepsilon_{UB}=\frac{1}{2}\frac{g^{2}}{\mu^{2}}\rho^{2}.$
For the purposes of the present study, it is sufficient to consider
neutron stars as simply consist of neutrons (n), protons (p) and
electrons (e). Including the Yukawa term the total pressure inside
neutron stars is $P=P_ {_{\textrm{npe}}}+P {_{\textrm{UB}}}$. For
the inner and outer crusts we use for $P_ {_{\textrm{npe}}}$ the EOS
of Carriere et al. \cite{Hor03} and that of Baym et al. \cite{BPS},
respectively. They are smoothly connected to the EOS in the core
\cite{Xu09}. For the latter we use $ P_{\rm
npe}(\rho,\delta)=\rho^2\left[dE_0/d\rho+dE_{\rm
sym}/d\rho\delta^2\right] +\frac{1}{2}\delta(1-\delta)\rho E_{\rm
sym}(\rho) $. The value of the isospin asymmetry $\delta$ at $\beta$
equilibrium is determined by the chemical equilibrium condition
$\mu_e=\mu_n-\mu_p=4\delta E_{\rm sym}(\rho)$ and the charge
neutrality requirement $\rho_e=\frac{1}{2}(1-\delta)\rho$. The
$E_0(\rho)$ and \esym obtained consistently within the modified GHF
approximation are \cite{Das03,Xu09}, respectively,
\begin{widetext}
\begin{eqnarray}
E_0(\rho)&=& \frac{8 \pi}{5 m h^3 \rho} p^5_f + \frac{\rho}{4
\rho_0}\cdot(-216.55)+ \frac{B}{\sigma + 1}
\left(\frac{\rho}{\rho_0}\right)^\sigma + \frac{1}{3 \rho_0 \rho}
(C_l + C_u) \left(\frac{4 \pi}{h^3}\right)^2 \Lambda^2
\notag\\\nonumber &\times& \left[p^2_f (6 p^2_f - \Lambda^2) - 8
\Lambda p^3_f \arctan \frac{2 p_f}{\Lambda}+\frac{1}{4} (\Lambda^4 +
12 \Lambda^2 p^2_f) \ln \frac{4 p^2_f +
\Lambda^2}{\Lambda^2}\right],
\end{eqnarray}
\begin{eqnarray}\label{esymmdi}
E_{sym}(\rho)&=& \frac{8 \pi}{9 m h^3 \rho} p^5_f + \frac{\rho}{4
\rho_0} [-24.59+4Bx/(\sigma +1)] - \frac{B x}{\sigma + 1}
\left(\frac{\rho}{\rho_0}\right)^\sigma \notag\\
&+& \frac{C_l}{9 \rho_0 \rho} \left(\frac{4 \pi}{h^3}\right)^2
\Lambda^2 \left[4 p^4_f - \Lambda^2 p^2_f \ln \frac{4 p^2_f +
\Lambda^2}{\Lambda^2}\right] + \frac{C_u}{9 \rho_0 \rho}
\left(\frac{4 \pi}{h^3}\right)^2 \Lambda^2 \left[4 p^4_f - p^2_f (4
p^2_f + \Lambda^2) \ln \frac{4 p^2_f + \Lambda^2}{\Lambda^2}\right],
\end{eqnarray}
\end{widetext}
where $p_f=\hbar(3\pi^2\frac{\rho}{2})^{1/3}$ is the Fermi momentum
for symmetric nuclear matter at density $\rho$. The coefficients
$A_{u}(x)=-95.98-x\frac{2B}{\sigma+1}$ and
$A_{l}(x)=-120.57+x\frac{2B}{\sigma +1}$. The values of the
parameters are $\sigma=4/3$, $B=106.35$ MeV, $C_{l}=-11.70$ MeV,
$C_{u}=-103.40$ MeV and $\Lambda=p_f^{0}\equiv
p_f(\rho_0)$~\citep{Das03}. The resulting symmetric EOS contribution
$dE_0/d\rho$ to the pressure is consistent with that extracted from
studying kaon production and nuclear collective flow in relativistic
heavy-ion collisions using hadronic transport models assuming no
hadron to Quark-Gluon Plasma phase transition up to about $5\rho_0$
\cite{Dan02a,LCK08}. The parameter $x$ in Eq. \ref{esymmdi} was
introduced to vary the density dependence of the \esym without
changing any property of symmetric nuclear matter and the value of
$E_{sym}(\rho_0)=31$ MeV \cite{Das03}. Shown in the inset of Fig.1
are two typical \esym denoted as MDIx1 and MDIx0 obtained by using
$x=1$ and $x=0$, respectively. While the MDIx0 \esym increases
continuously, the MDIx1 \esym becomes negative above $3\rho_0$. Only
the MDIx1 \esym can reproduce the FOPI/GSI pion production data
within the transport model analysis \cite{Xiao09}. It is seen that
the corresponding MDIx1 pressure decreases with increasing density
as shown with the lowest curve in Fig.\ 1. However, the Yukawa term
makes the pressure grow continuously with increasing density with a
value of $g^2/\mu^2$ higher than about 10 GeV$^{-2}$.

\begin{figure}
\includegraphics[width=0.4\textwidth]{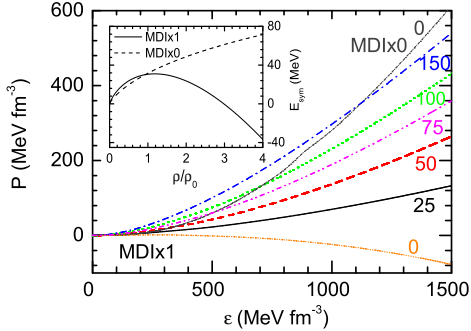}
\vspace{-0.7 cm} \caption{\label{fig1}  (Color online)The inset
shows two typical examples (MDIx0 and MDIx1) of the density
dependence of the nuclear symmetry energy. The MDIx1 (MDIx0) EOS
with (without) the Yukawa contribution using different values of the
$g^{2}/\mu^{2}$ in units of GeV$^{-2}$ are shown.} \vspace{-0.7cm}
\end{figure}

\begin{figure}
\includegraphics[width=0.4\textwidth]{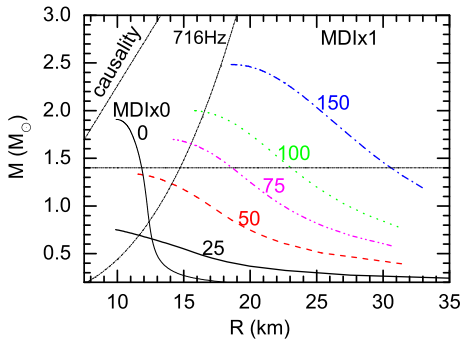}
\vspace{-0.4 cm} \caption{\label{fig2}  (Color online) The
mass-radius relation of static neutron stars with the MDIx1 (MDIx0)
EOS with (without) the Yukawa contribution. The static neutron star
sequences constrained by the rotational frequency 716 Hz of the
J1748-2446ad \cite{Hessels06} are taken from Haensel et al.
\cite{Haensel09}.} \vspace{-0.5cm}
\end{figure}

Shown in Fig.\ \ref{fig2} is the mass-radius relation of static
neutron stars obtained from solving the Tolman-Oppenheimer-Volkoff
(TOV) equation using the MDIx1 \esym and various values for the
$g^{2}/\mu^{2}$. The result obtained using the MDIx0 without
including the Yukawa term is included as a reference
\cite{LiBA06}. The causality \cite{Lat04} and rotational
constraint \cite{Haensel09} are also shown. The Keplerian
(mass-shedding) frequency is approximately \cite{Haensel09} $
\nu_{ k}\approx 1.08 \Big(\frac{M}{
\textrm{M}_{\odot}}\Big)^{1/2}\Big(\frac{R}{10~\textrm{km}}\Big)^{-3/2}~
\textrm{kHz}. $ So far, the fastest pulsar observed is the
J1748-2446ad spinning at 716 Hz \cite{Hessels06}. Taking 716 Hz as
the Keplerian frequency, the M-R relation is restricted to the
left side of the rotational limit. The latter restricts the value
of $g^{2}/\mu^{2}$ to less than 150 GeV$^{-2}$. It is seen that to
produce a neutron star with a maximum mass above $1.4
~\textrm{M}_{\odot}$, the $g^{2}/\mu^{2}$ has to be higher than
about 50 GeV$^{-2}$. More specifically, with the MDIx1 \esym and
the $g^{2}/\mu^{2}=50-150$ GeV$^{-2}$, or equivalently
$|\alpha|\lambda^2=(2.6-7.8)\times 10^{7} \rm{m}^2$, neutron stars
can have a maximum mass between 1.4 and 2.5 $\textrm{M}_{\odot}$
and a corresponding radius between 12 and 18 km.

For canonical neutron stars of 1.4 $\textrm{M}_{\odot}$, the radius
is quite sensitive to the $g^{2}/\mu^{2}$ value used. Thus, besides
the accurate measurement of neutron star radii, additional
measurements related to the mass distribution, such as the moment of
inertia, will be very useful in setting astrophysical constraints on
the \esym and $g^{2}/\mu^{2}$. According to Lattimer and Schutz
\cite{Lattimer05}, at the slow rotation limit the moment of inertia
can be well approximated as
$
I\approx (0.237\pm 0.008)MR^{2}\Big[1+4.2\frac{M
}{\textrm{M}_{\odot} }\cdot\frac{ \textrm{km}}{ R}+90\Big(\frac{M
}{\textrm{M}_{\odot} }\cdot\frac{ \textrm{km}}{ R}\Big)^{4}\Big].
$
\begin{figure}
\includegraphics[width=0.4\textwidth]{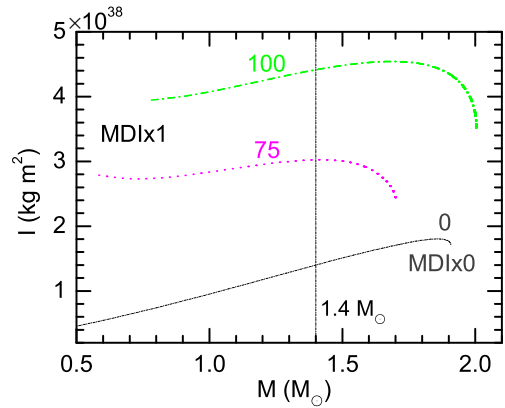}
\vspace{-0.3 cm} \caption{\label{fig3}  (Color online) The momenta
of inertia of neutron stars with the MDIx1 (MDIx0) \esym with
(without) the Yukawa contribution. The numbers above the lines are
the $g^{2}/\mu^{2}$ values in units of GeV$^{-2}$.} \vspace{-0.45cm}
\end{figure}
Shown in Fig. 3 is the $I$ as a function of M. For
$M=1.4~\textrm{M}_{\odot}$, the MDIx0 without the Yukawa
contribution gives an $I$ no more than $1.8\times
10^{38}~\textrm{kg}\cdot \textrm{m}^{2}$ \cite{Worley08}. However,
significantly larger $I$ values are obtained with the MDIx1 \esym
and the Yukawa contribution. The discovery of the double-pulsar
system PSR J0737-3039 $A\&B$ provides a great opportunity to
determine accurately the moment of inertia $I_{A}$ of the star $A$
\cite{Lyne04, Lattimer07}. Our results shown here add to the
importance of measuring the moment of inertia accurately.

To constrain the values of $\alpha$ and $\lambda$ has been a
longstanding goal of many terrestrial experiments and
astrophysical observations as limits on them may provide useful
guidance for developing grand unification theories, see, e.g.,
refs.
\cite{Fis99,Bordag01,Adel03,Fujii71,Gib81,Ser05,Decca05,Ade07,Kap07,Kamyshkov08}.
These studies have estimated various upper limits on the $\alpha$.
In the range of $\alpha=10^{-10}-10^{38}$ and
$\lambda=10^{15}-10^{-14}$ m, there is a clear trend of increased
strength $\alpha$ at shorter length $\lambda$. What we have
constrained is the value of $g^{2}/\mu^{2}$ or equivalently the
$|\alpha|\lambda^2$ from the pressure necessary to support both
static neutron stars and the fastest pulsars. While we expect that
the range parameter $\lambda$ has to be much larger (smaller) than
the radii of finite nuclei (neutron stars), we can not set
separate constraints on the values of $\alpha$ and $\lambda$.
Compared to other efforts to constrain the $\alpha$ and $\lambda$,
our study here is unique in that the estimated minimum value of
$g^{2}/\mu^{2}$ is a lower limit satisfying all known constraints
from both terrestrial nuclear experiments and observations of
global properties of neutron stars. Moreover, very interestingly,
our estimated range of $g^{2}/\mu^{2}$ overlaps well with the
upper limits estimated from analyzing the neutron-proton and
neutron-lead scattering data in the range of $\lambda \approx
10^{-14}-10^{-8}$ m \cite{Kamyshkov08,BARB75,POKO06,NESV08}.

In summary, neutron stars are a natural testing ground of grand
unification theories of fundamental forces. Considering the possible
violation of the ISL of gravity, the stability and observed
properties of NSs can not rule out super-soft symmetry energies at
supra-saturation densities. Given the uncertainties and model
dependence involved in extracting information about the EOS and
symmetry energy from heavy-ion reactions, it is very important to
test the possible super-soft symmetry energy at supra-saturation
densities using several observables simultaneously from independent
experiments analyzed using different models. If confirmed, it may
point towards a violation of the ISL in neutron stars.

We would like to thank M. I. Krivoruchenko for useful communications
and G. C. Yong, C. Xu and J. Xu for helpful discussions. The work is
supported in part by the National Natural Science Foundation of
China under Grant No. 10647116, 10710172, 10575119, 10675082 and
10975097, the Young Teachers' Training Program from China
Scholarship Council under Grant No. 2007109651, the MOE of China
under project NCET-05-0392, Shanghai Rising-Star Program under Grant
No.06QA14024, the SRF for ROCS, SEM of China, and the National Basic
Research Program of China (973 Program) under Contract
No.2007CB815004, the US National Science Foundation under Grants No.
PHY0652548 and No. PHY0757839, the Research Corporation under Award
No. 7123 and the Texas Coordinating Board of Higher Education Grant
No.003565-0004-2007.


\begin{thebibliography}{1}

\bibitem{Sum94}K. Sumiyoshi and H. Toki, ApJ, {\bf 422}, 700 (1994).

\bibitem{Bom01}I. Bombaci, Ch. 2 in Isospin Physics in Heavy-Ion Collisions at Intermediate
Energies, Eds. Bao-An Li and W. Udo Schr\"{o}der (Nova Science
Publishers, Inc, New York, 2001).

\bibitem{Lat04} J.M. Lattimer, M. Prakash, ApJ, {\bf 550}, 426 (2001); Science {\bf 304}, 536 (2004).

\bibitem{Ste05a} A.W. Steiner et al., Phys. Rep. {\bf 411}, 325
(2005).

\bibitem{LiBA98} B.A. Li et al., Int. Jour. Mod. Phys. E {\bf 7}, 147 (1998).

\bibitem{Dan02a} P. Danielewicz et al., Science {\bf 298}, 1592 (2002).

\bibitem{Bar05} V. Baran et al., Phys. Rep. {\bf 410}, 335 (2005).

\bibitem{LCK08} B.A. Li et al., Phys. Rep. {\bf 464}, 113 (2008).

\bibitem{Bro00} B.A. Brown, Phys. Rev. Lett. {\bf 85}, 5296 (2000).

\bibitem{Kut94} M. Kutschera, Phys. Lett. {\bf B340}, 1 (1994).

\bibitem{Kub99}S. Kubis and M. Kutschera, Nucl. Phys. {\bf A720},
189 (2003).

\bibitem{Szm06} A. Szmaglinski et al., Acta Phys. Polon. {\bf
B37}, 277 (2006).

\bibitem{Dip03} A.E. L. Dieperink et al., Phys. Rev. C{\bf 68}, 064307
(2003).
\bibitem{Lee98}C.-H. Lee et al., Phys. Rev. C{\bf 57}, 3488 (1998).

\bibitem{Hor01a} C.J. Horowitz et al., Phys. Rev. Lett {\bf 86}, 5647 (2001).

\bibitem{Che07} L.W. Chen et al., Phys. Rev. C \textbf{76},
054316 (2007).

\bibitem{LiZH06} Z.H. Li et al., Phys. Rev. C \textbf{74}, 047304 (2006).

\bibitem{Pan72} V.R. Pandharipande, et al., Phys. Lett. {\bf B39}, 608 (1972).

\bibitem{Fri81} B. Friedman et al., Nucl. Phys. {\bf A361}, 502 (1981).

\bibitem{Wir88a} R.B. Wiringa et al., Phys. Rev. {\bf C38}, 1010 (1988).
\bibitem{eft} N. Kaiser et al., Nucl. Phys. {\bf A697}, 255
(2002).

\bibitem{Kra06} P. Krastev et al., Phys. Rev. {\bf C74}, 025808 (2006).

\bibitem{Cha97}E. Chabanat et al., Nucl. Phys. A 627 (1997) 710;
{\it ibid}, 635 (1998) 231.

\bibitem{Sto03} J.R. Stone et al., Phys. Rev. {\bf C68}, 034324 (2003).

\bibitem{Che05b} L.W. Chen et al., Phys. Rev. C \textbf{72},
064309 (2005).

\bibitem{Dec80}J. Decharge et al., Phys. Rev. {\bf C21},
1568 (1980).

\bibitem{Das03}C. B. Das et al., Phys. Rev. {\bf C67}, 034611 (2003).

\bibitem{MS}W.D. Myers et al., Acta Phys. Pol. {\bf B26},111 (1995).

\bibitem{Kho96} D.T. Khoa et al., Nucl. Phys. {\bf A602}, 98 (1996).

\bibitem{Bas07} D.N. Basu et al, Nucl. Phys. {\bf A811}, 140 (2008).

\bibitem{Ban00}S. Banik et al., J. Phys. G {\bf 26}, 1495 (2000).

\bibitem{Ch09}P. Roy Chowdhury et al., Phys. Rev. C{\bf 80}, 011305
(R) (2009).

\bibitem{Xiao09}Z.G. Xiao et al., Phys. Rev. Lett. {\bf 102}, 062502 (2009).

\bibitem{Rei07} W. Reisdorf et al., Nucl. Phys. A {\bf 781}, 459 (2007).

\bibitem{IBUU04} B.A. Li et al., Nucl. Phys. {\bf A735}, 563 (2004).

\bibitem{Glen}N. Glendening, Compact Stars, Springer, New York (2000), ISBN
0387989773.

\bibitem{Stone05} J. R. Stone, J. Phys. G{\bf 31},
R211 (2005).

\bibitem{Pea01}R. Pease, Nature (London) {\bf 411}, 986 (2001).

\bibitem{Hoy03}C.D. Hoyle, Nature (London) {\bf 421}, 899 (2003).

\bibitem{Ark98} N. Arkani-Hamed et al., Phys Lett. {\bf
B429}, 263 (1998). 

\bibitem{Fis99}E. Fischbach and C.L. Talmadge, The Search for
Non-Newtonian Gravity, Springer-Verlag, New York, Inc. (1999), ISBN
0-387-98490-9.

\bibitem{Adel03}E. G. Adelberger et al., Annu. Rev. Nucl. Part. Sci. \textbf{53}, 77 (2003).

\bibitem{Fujii71}Y. Fujii, Nature (London) {\bf 234}, 5 (1971).

\bibitem{Long03}J.C. Long et al., Nature (London) {\bf 421}, 922 (2003).

\bibitem{Uzan03}J.P. Uzan, Rev. Mod. Phys., {\bf 75}, 403 (2003).

\bibitem{Kri09} M.I. Krivoruchenko et al., Phys. Rev. {\bf D79}, 125023 (2009).

\bibitem{Fayet}P. Fayet, Phys. Lett. {\bf B675}, 267 (2009).

\bibitem{Boe04}C. Boehm and P. Fayet, Nucl. Phys. {\bf B683}, 219 (2004).

\bibitem{Jean03} P. Jean et al., \textit{et al.}, A\&A \textbf{407}, L55 (2003).
\bibitem{Boehm04a} C. Boehm et al., Phys. Rev. Lett.   \textbf{92}, 101301 (2004); Phys. Rev. D \textbf{69}, 101302 (2004).

\bibitem{Zhu07}S.H. Zhou, Phys. Rev. D \textbf{75}, 115004 (2007).

\bibitem{Fuj2}Y. Fujii, in Large Scale Structures of the Universe,
page 471-477, Eds. J. Audouze et al. (1988), International
Astronomical Union.

\bibitem{Hor03} J. Carriere et al., ApJ, {\bf 593}, 463 (2003).

\bibitem{BPS} G. Baym et al., ApJ, {\bf 170}, 299 (1971).

\bibitem{Xu09} J. Xu et al., ApJ. {\bf 697}, 1549
(2009).

\bibitem{Hessels06} J. W. T. Hessels \textit{et al.},  Science  \textbf{311}, 1901
(2006).

\bibitem{Haensel09} P. Haensel et al., A\&A 502, 605 (2009).

\bibitem{LiBA06}B.A. Li et al., Phys. Lett. {\bf B642}, 436 (2006).

\bibitem{Lattimer05} J. M. Lattimer et al.,  Astrophys. J. \textbf{629}, 979
(2005).

\bibitem{Worley08}A, Worley et al., ApJ \textbf{685}, 390
(2008).

\bibitem{Krastev08} P, G. Krastev et al., ApJ \textbf{676}, 1170
(2008).

\bibitem{Lyne04}A.G. Lyne et al., Science \textbf{303}, 1153 (2004).

\bibitem{Lattimer07} J. M. Lattimer et al., Phys. Rep. \textbf{442}, 109
(2007).

\bibitem{Gib81}G.W. Gibbons and B.F. Whiting, Nature (London)
{\bf 291}, 636 (1981).

\bibitem{Bordag01}M. Bordag et al., Phys. Rep. \textbf{353}, 1 (2001).

\bibitem{Ser05} Serge Reynaud et al., Int. J. Mod. Phys. {\bf A20},  2294
(2005).

\bibitem{Decca05}R. S. Decca et al., Phys. Rev. Lett. \textbf{94}, 240401 (2005).

\bibitem{Ade07} E.G. Adelberger et al., Phys. Rev. Lett. {\bf 98}, 131104
(2007).

\bibitem{Kap07} D.J. Kapner et al., Phys. Rev. Lett. {\bf 98},  021101
(2007).

\bibitem{Kamyshkov08}Y. Kamyshkov et al., Phys. Rev. D{\bf 78}, 114029 (2008).

\bibitem{BARB75} R. Barbieri et al., Phys. Lett. B{\bf 57}, 270 (1975).

\bibitem{POKO06}Yu.N. Pokotilovski, Phys.\ Atom.\ Nucl.\  {\bf 69}, 924 (2006).

\bibitem{NESV08} V.V. Nesvizhevsky \textit{et al.}, Phys. Rev. D{\bf 77}, 034020 (2008).

\end{thebibliography}
\end{document}